\documentclass{article}
\usepackage{spconf,amsmath,graphicx,hyperref,booktabs,arydshln,amssymb,multirow,float,xcolor,nccmath}

\title{ARCHI-TTS: A flow-matching-based Text-to-Speech Model with Self-supervised Semantic Aligner and Accelerated Inference}
\name{Chunyat Wu, Jiajun Deng, Zhengxi Liu, Zheqi Dai, Haolin He, Qiuqiang Kong}
\address{
  The Chinese University of Hong Kong, Hong Kong SAR, China}
\begin{document}
\ninept
\maketitle

\begin{abstract}
Although diffusion-based, non-autoregressive text-to-speech (TTS) systems have demonstrated impressive zero-shot synthesis capabilities, their efficacy is still hindered by two key challenges: the difficulty of text-speech alignment modeling and the high computational overhead of the iterative denoising process.
To address these limitations, we propose ARCHI-TTS that features a dedicated semantic aligner to ensure robust temporal and semantic consistency between text and audio.  To overcome high computational inference costs, ARCHI-TTS employs an efficient inference strategy that reuses encoder features across denoising steps, drastically accelerating synthesis without performance degradation. An auxiliary CTC loss applied to the condition encoder further enhances the semantic understanding. 
Experimental results demonstrate that ARCHI-TTS achieves a WER of 1.98\% on LibriSpeech-PC test-clean, and 1.47\%/1.42\% on SeedTTS test-en/test-zh with a high inference efficiency, consistently outperforming recent state-of-the-art TTS systems.
Audio samples and code are publicly available at \url{https://archimickey.github.io/architts}.

\end{abstract}

\section{Introduction}
\label{sec:intro}

Text-to-Speech (TTS) synthesis aims at generating speech audios with a given text prompt and has witnessed remarkable advancements in recent years \cite{zeng2024glm,chen2025f5,du2024cosyvoice2}. Conditioned on a few seconds of an audio prompt, current TTS models can perform zero-shot synthesis \cite{du2024cosyvoice2, chen2025f5}, generating high-fidelity, natural speech for arbitrary text while accurately mimicking the speaker's distinct vocal characteristics.

Current zero-shot TTS are broadly spearhead into two main categories: autoregressive (AR) models and non-autoregressive (NAR) models. \textbf{AR-based TTS models} consecutively predict discrete audio tokens conditioned on text and previous audio tokens, exhibit promising zero-shot TTS capabilities \cite{du2024cosyvoice, anastassiou2024seed, liu2025e1}. Despite their impressive performance, which is often enhanced by implicit duration modeling and diverse sampling strategies, their foundational AR structure presents inherent limitations, for example, inference latency, exposure bias, as well as a strong dependency on the quality of the speech tokenizer \cite{zeng2024glm, du2024cosyvoice} for high-fidelity synthesis. In contrast, \textbf{NAR-based TTS models} have gained prominence by offering parallel synthesis, which can effectively address the limitations of AR modeling. A key catalyst for recent breakthroughs in NAR speech synthesis is Flow Matching with Optimal Transport Path, which directly models continuous data distributions, for example, NaturalSpeech3 \cite{ju2024naturalspeech} and F5-TTS \cite{chen2025f5}. 

Efforts on NAR TTS models are primarily concentrated on two crucial challenges: \textbf{a)} the effective text-speech alignment modeling, and \textbf{b)} the computational intensity mitigating of iterative denoising. Regarding text-speech alignment, researchers have explored various strategies. Several approaches relied on explicit guidance, such as phoneme-level duration utilized in NaturalSpeech3 \cite{ju2024naturalspeech} and Voicebox \cite{le2023voicebox}, or monotonic alignment search with an auxiliary duration predictor utilized in Matcha-TTS \cite{mehta2024matcha}. However, recent studies suggest that such rigid alignment information can compromise the naturalness of the synthesized speech \cite{eskimez2024e2}. Consequently, simpler yet highly effective methods have gained traction, such as padding character tokens to match the length of the speech representation, a technique used by E2-TTS \cite{eskimez2024e2} and F5-TTS \cite{chen2025f5} to achieve remarkably natural and realistic speech synthesis. To mitigate the high computational cost of iterative denoising, research has focused on efficient diffusion sampling, with Diffusion Model Distillation (DMD) emerging as a popular strategy for reducing the Number of Function Evaluations (NFE), for example, teacher-student distillation utilized in DMDSpeech \cite{li2024dmdspeech} and adversarial distillation utilized in E1-TTS \cite{liu2025e1} are proposed. However, a significant drawback of these distillation methods is the increased training overhead. They necessitate not only a pre-trained teacher model but also extra forward passes within the training loop, leading to a more complicated training. 

To address the aforementioned two challenges, this paper propose a NAR \textbf{archi}tecturally refined \textbf{text-to-speech} (ARCHI-TTS) model for high-quality and fast speech synthesis. ARCHI-TTS is a novel aligner-encoder-decoder architecture trained with a flow-matching objective. First, to model the text-speech alignment, a semantic aligner is proposed to produce self-supervised text-aligned semantic representations in flexible length through end-to-end training with flow-matching based decoder. Second, to improve efficiency, ARCHI-TTS enables fast generation by reusing the condition encoder output across multiple decoding steps without extra distillation, significantly reducing the computational intensity of iterative denoising. 
Experiments conducted on the large-scale 100k-hour multi-lingual Emilia dataset \cite{he2024emilia} demonstrate several key finding: \textbf{a) Competitive performance with high efficiency:} ARCHI-TTS outperforms state-of-the-art (SOTA) TTS models on the LibriSpeech test-clean and SeedTTS test set, despite being trained with a fraction of the typical computational resources (8 RTX5090 GPUs for 4 days) and training data. \textbf{b) Training-free inference acceleration:} As a direct benefit of a by-product of separated DiT architecture, the model can be significantly accelerated at inference time without a significant degradation in synthesized speech quality. \textbf{c) Competitive subjective quality:} In mean opinion score (MOS) evaluations, ARCHI-TTS achieves ratings for naturalness, speaker similarity, and quality that are competitive with large-scale, industrial TTS systems.

This paper is organized as follows: Section \ref{sec:method} introduces the architecture of ARCHI-TTS and its main components. Section \ref{sec:exp} describes the experimental setup, including datasets, model configurations, training details, and evaluation metrics. Section \ref{sec:results} presents the results and analysis of our experiments
Section \ref{sec:conclusion} concludes the paper and discusses future work.

\section{ARCHI-TTS}
\label{sec:method}
\begin{figure}[!t]
  \centering
  \includegraphics[width=0.85\linewidth]{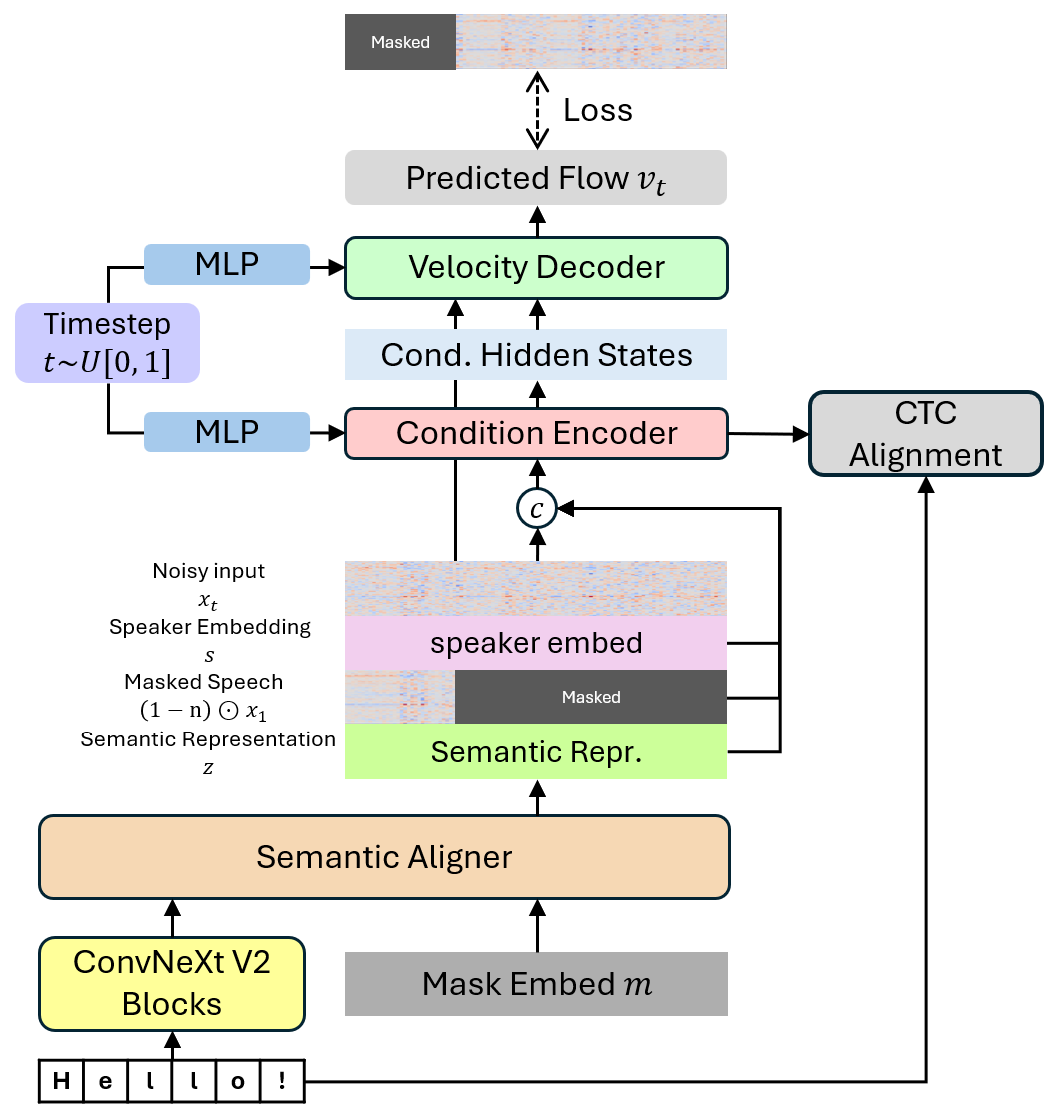}
  \vspace{-0.3cm}
  \caption{The overview of the ARCHI-TTS architecture.}
  \label{fig:architecture}
  \vspace{-0.5cm}
\end{figure}

ARCHI-TTS is a fully non-autoregressive speech synthesis model that generates speech conditioned on both a text input and a short audio prompt. The architecture is centered on two primary components: a semantic aligner and a flow matching decoder. The semantic aligner is tasked with creating an alignment between the text embeddings and the corresponding speech representations. The core of the synthesis process is the flow matching decoder, which is constructed from Diffusion Transformer (DiT) blocks. The overall architecture is illustrated in Figure \ref{fig:architecture},  

\subsection{Semantic Aligner}
\label{ssec:aligner}

A key challenge in NAR TTS synthesis is reconciling the potential length mismatch between the input text tokens and the output speech frames. Recent studies \cite{zeng2024glm, du2024cosyvoice2, anastassiou2024seed} have shown that the full attention mechanism within a transformer is exceptionally effective at capturing complex relationships between different modalities. Therefore, to explicitly learn the alignment between the semantic content of the text and the temporal structure of the speech, we introduce a semantic aligner built upon a transformer encoder.
The aligner operates on two primary input sequences: \textbf{a) Text Sequence:} The input text, represented as a sequence of character or pinyin tokens, $y$, is first encoded into embeddings. These embeddings are then passed through several ConvNeXt V2 \cite{woo2023convnext} blocks to produce a sequence of rich semantic features. \textbf{b) Speech-Length Sequence:} To represent the temporal duration of the target speech, a masked sequence is created by replicating a learnable mask embedding, $m$, for $N$ times, where $N$ corresponds to the length of the speech latent. Before being processed by the transformer, a learnable start-of-sequence token is prepended to each sequence. The entire operation of the semantic aligner can be formulated as:
\begin{equation}
z = \textbf{Transformer}(e_{st}, y, e_{sm}, \underbrace{m, \ldots, m}_{N \text{times}}),
\end{equation}
where $e_{st}$ and $e_{sm}$ are learnable special tokens representing the start of the text and mask sequences, respectively. Conceptually, the replicated mask embeddings serve as a uniform temporal canvas, initially encoding only the duration of the target speech. The semantic aligner then leverages its transformer blocks to aggregate the semantic representations from the text features and align them with this temporal canvas, which effectively converts a simple sequence of duration markers into a rich sequence of contextually-aware semantic features. Another benefit of this mechanism is its ability to remove the dependency of the speech audio length on the text token length. This is a critical advantage, as we observe that it is possible for text tokens to be shorter than the speech audio frames, especially under low-token-rate scenarios that arise from character tokenization.

\subsection{Compressed Speech Latent Representation}
\label{ssec:vae}
Conventional NAR TTS pipelines often rely on mel-spectrograms as the intermediate acoustic representation. A primary limitation of this approach is the high temporal redundancy of mel-spectrograms, which results in a high token rate (e.g., 50-100Hz) and necessitates a separate, complex vocoder model to invert the representation back to an audio waveform.
To address these challenges, our work adopts a more direct synthesis approach by utilizing a highly compressed, low-token-rate latent representation derived from a Variational Autoencoder (VAE). The adoption of a VAE is motivated by its demonstrated success as a neural audio compressor, particularly when its latent space is regularized with a Kullback--Leibler (KL) divergence objective \cite{rombach2022high}. This allows the VAE to function as both an encoder and a decoder, unifying the representation and synthesis stages. Following the architectural principles of Stable Audio \cite{evans2025stable}, we trained a custom VAE to encode 24kHz speech signals into a sequence of continuous latents at a token rate of 12.5Hz.

\subsection{Condition Encoder with Flow Matching}
The speech decoder in ARCHI-TTS is responsible for generating the VAE latent representation conditioned on various inputs. To accomplish this, we leverage the conditional flow matching (CFM) framework \cite{lipman2023flow}, a powerful approach for generative modeling. CFM models are designed to learn a time-dependent vector field, $v_t(x_t; \theta)$, which generates a flow $\phi_t$ capable of transforming a simple prior distribution $p_0$ (e.g., a standard Gaussian) into a target data distribution $p_1$ that matches the true data distribution $q$. Following the optimal transport path, the trajectory between a noise sample $x_0 \sim p_0$ and a data sample $x_1 \sim q$ is defined as a linear interpolation: $x_t = (1 - t)x_0 + t x_1$, for $t \in [0, 1]$. To parameterize the conditional vector field $v_t$, we adapt DiT to a novel encoder-decoder architecture, comprising a condition encoder and a velocity decoder \cite{wang2025ddt}. The model is conditioned on a rich set of inputs: \textbf{a) semantic features} $z$ provided by the semantic aligner to guide the content of the speech; \textbf{b) speaker embedding} $s$ that is a global embedding, replicated to match the speech length, to control the coarse-grained speaker timbre;
and \textbf{c) audio prompt} $x_{\text{ref}}$ that is a masked segment of the target speech latent, designed to preserve fine-grained speaker timbre. It is derived by applying a binary mask $n$ with a random start position to the ground-truth latent, denoted as $x_{\text{ref}} = (1 - n) \odot x_1$.

The condition encoder then generates conditioned hidden states $h$, which provide rich contextual information for the velocity decoder to predict the flow velocity $v_t$. The CFM loss which encourages the predicted velocity to match the ground-truth velocity of the optimal transport path is expressed as:
\begin{equation}
\label{eq:cfmloss2}
    {\cal L}_{\text{CFM}} = \mathbb{E}_{t,\, q(x_1),\, p_0(x_0)} \left[ \| v_t(x_t, x_{\text{ref}}, z, s; \theta) - \hat{v}_t \|^2 \right],
\end{equation}

To further enhance the semantic alignment between the hidden representation and the input text, we introduce an auxiliary text alignment loss based on the Connectionist Temporal Classification (CTC) framework, which is expressed as  
\begin{equation}
\label{eq:text_loss1}
    {\cal L}_{\text{CTC}} = -\log p_{\text{CTC}} \left( y \mid \Phi_i \left( v_t(x_t, x_{\text{ref}}, z, s;\theta) \right) \right),
\end{equation}
where $\Phi_i$ denotes the intermediate hidden representation from the $i$-th DiT block of the condition encoder.

\subsection{Velocity Decoder}
\label{ssec:decoder}
The goal of velocity decoder is to predict the velocity vector field $v_t$ for each step $t$ of the denosing process. Architecturally, it is composed of a series of DiT blocks and a final projection layer that outputs the predicted velocity. The velocity decoder takes two primary inputs: the noisy speech latent $x_t$ and the conditioned hidden states $h$ from the condition encoder. Instead of simply concatenating $h$ with $x_t$, which would treat the conditioning signal as a local feature, we inject $h$ as a global condition. Specifically, $h$ is added to the sinusoidal timestep embedding, and this combined embedding is fed into each DiT block. 
To accelerate convergence and improve synthesis quality, we enhance the training objective by incorporating a velocity direction loss ${\cal L}_{\text{DIR}}$ and employing logit-normal timestep sampling. The direction loss uses cosine similarity to ensure the correct flow orientation, while logit-normal sampling focuses training on the more challenging start and end points of the generation trajectory. The final training objective is expressed as 
\begin{equation}
    {\cal L} = {\cal L}_{\text{CFM}} + {\cal L}_{\text{DIR}} + \eta \cdot {\cal L}_{\text{CTC}}
\end{equation}
where $\eta$ is a hyperparameter.

\subsection{Zero-shot TTS}
\label{ssec:inference}
The zero-shot speech synthesis generates the speech audio corresponding to a target text $y_{\text{ref}}$, conditioned on the reference audio $x_{\text{ref}}$, the reference transcription $y_{\text{ref}}$, and the speaker embedding $s_{\text{ref}}$. The duration of the output VAE latent sequence is estimated by preserving the token-per-frame rate of the reference audio, ensuring prosodic consistency. Specifically, given $T_{\text{ref}}$ latent frames for the reference audio and $L_{\text{ref}}$ and $L_{\text{gen}}$ text tokens for the reference and target transcriptions respectively, the target duration is calculated as $d = \left\lfloor L_{\text{gen}} \cdot (T_{\text{ref}}/L_{\text{ref}}) \right\rfloor$. For the semantic conditioning, a unified context by concatenating the reference and target transcriptions ($y_{\text{ref}}$ and $y_{\text{gen}}$) is used to extract a single semantic feature vector, $z_{\text{ref} \cdot \text{gen}}$, through the semantic aligner to guide the synthesis. Finally, the Euler solver is employed to generate the final VAE latent representation through the condition encoder and velocity decoder, which is then transformed back into speech audio using the VAE decoder.
\subsection{Sampling}
\label{ssec:inference}
Sample generation is performed by solving the ODE defined by the learned velocity field, transforming noise into a data sample. To enhance quality, we employ Classifier-Free Guidance (CFG) \cite{ho2021classifier}. During inference, the guided velocity is computed by extrapolating from the unconditional prediction towards the conditional one,
\begin{equation}
    \tilde{v}_t(x_t, c, \omega; \theta) = (1 + \omega)v_t(x_t, c; \theta) - \omega v_t(x_t, \emptyset; \theta)
\label{eq:cfg}
\end{equation}
where $c$ and $\emptyset$ denote the condition and the zero (unconditional) condition, respectively, and $\omega$ is the CFG strength parameter that controls the trade-off between fidelity and diversity. During inference, intermediate conditioned hidden states $h_t$ can be shared across adjacent $t$ to achieve accelerated sampling. Given $N$ NFE steps and $K$ sharing steps, we denote the sharing ratio as $1 - K/N$. This mechanism allows us to bypass the encoder component of the model—which typically dominates the computational cost of the forward pass—by reusing the stored encoder outputs from previous timesteps, thereby enabling accelerated inference.

\vspace{-1.5em}
\section{Experimental Setup}
\vspace{-1em}
\label{sec:exp}

\begin{table}[!t]
    \centering
    \caption{Results on LibriSpeech-PC test-clean. RTF results are measured on a single RTX 3090 GPU inferencing a 10-second audio. * denotes results from the original papers. $\diamondsuit$ denotes results from F5-TTS paper. $\dagger$ denotes the results from DiTAR paper.}
    \resizebox{\linewidth}{!}{
    \begin{tabular}{lrcccc}
        \toprule
        \bf Model & \#Param. & \#Data & WER(\%)↓ & SSIM↑ & RTF↓ \\
        \midrule
        \multicolumn{6}{c}{\textbf{LibriSpeech-PC \textit{test-clean}}} \\
        \midrule
        \multicolumn{3}{l}{Ground Truth (\textit{1127 samples 2 hrs})}         &2.23          &0.69           &-  \\
        \multicolumn{3}{l}{VAE Resynthesized}                     &\textbf{2.19}          &\textbf{0.67}           &-  \\
        \multicolumn{3}{l}{Vocos \cite{siuzdak2023vocos} Resynthesized$^\diamondsuit$}                              &2.32          &0.66           &-  \\
        \cdashline{1-6}\noalign{\vskip\belowrulesep}
        CosyVoice \cite{du2024cosyvoice}$^\diamondsuit$ & $\sim$300M  &170K Multi. &3.59          &0.66           &0.92\\
        FireRedTTS \cite{guo2024fireredtts}$^\diamondsuit$ & $\sim$580M  &248K Multi. &2.69          &0.47           &0.84\\
        MaskGCT \cite{wang2024maskgct}* & $\sim$1.1B & 100K Multi. & 2.72 & 0.69 & - \\ 
        E2 TTS \cite{eskimez2024e2}$^\diamondsuit$  & 333M   &100K Multi. &2.95          &0.69  &0.68\\
        F5-TTS \cite{chen2025f5}$^\diamondsuit$ & 336M  &100K Multi. &2.42 &0.66           &0.31\\
        DiTAR \cite{jia2025ditar}$^\dagger$ & $\sim$600M   &100K Multi. &2.39 &0.67   &- \\
        \cdashline{1-6}\noalign{\vskip\belowrulesep}
        ARCHI-TTS & 289M & 100K Multi. & \textbf{1.98}         &\textbf{0.70}           &\textbf{0.21} \\
        \bottomrule
    \end{tabular}
    }
    \label{tab:librispeech}
    \vspace{-0.6cm}
\end{table}

\begin{table}[!t]
    \centering
    \caption{Results on Seed-TTS testset. * denotes results from the original papers. $\diamondsuit$ denotes results from F5-TTS's paper. $\dagger$ denotes results from DiTAR's paper.}
    \resizebox{\linewidth}{!}{
    \begin{tabular}{lcccc}
        \toprule
        \bf Model & \multicolumn{2}{c}{\bf Seed-EN} & \multicolumn{2}{c}{\bf Seed-ZH} \\
        \midrule
        & \bf WER(\%)↓ & \bf SSIM↑ & \bf WER(\%)↓ & \bf SSIM↑\\
        \midrule
        Ground Truth & 2.06 & 0.73 & 1.254 & 0.75 \\
        VAE Resynthesized & \textbf{2.06} & 0.70 & 1.44 & 0.71 \\
        Vocos \cite{siuzdak2023vocos} Resynthesized$^\diamondsuit$ & 2.09 & 0.70 & 1.27 & 0.72 \\
        \cdashline{1-5}\noalign{\vskip\belowrulesep}
        CosyVoice 2 \cite{du2024cosyvoice2}$^\dagger$ & 2.57 & 0.65 & 1.45 & 0.75 \\
        FireRedTTS\cite{guo2024fireredtts}$^\dagger$ & 3.82 & 0.46 & 1.51 & 0.63 \\
        MaskGCT\cite{wang2024maskgct}* & 2.623 & \underline{0.717} & 2.273 & 0.774 \\
        Seed-TTS$_{DiT}$ \cite{anastassiou2024seed}* & \textbf{1.733} & \textbf{0.790} & \textbf{1.178}  & \textbf{0.809} \\
        E2 TTS \cite{eskimez2024e2}$^\dagger$ & 2.19 & 0.71 & 1.97 & 0.73 \\
        F5-TTS \cite{chen2025f5}$^\dagger$ & 1.83 & 0.67 & 1.56 & 0.76 \\
        DiTAR \cite{jia2025ditar}$^\dagger$ & \textbf{1.69} & 0.74 & \textbf{1.02} & 0.75 \\
        \cdashline{1-5}\noalign{\vskip\belowrulesep}
        ARCHI-TTS & \textbf{1.47} & 0.68 & \underline{1.42} & 0.70 \\
        \bottomrule
    \end{tabular}
    }
    \label{tab:seedtts}
    \vspace{-0.6cm}
\end{table}

\subsection{Datasets}
\label{ssec:datasets}
For training, we primarily utilize the Emilia dataset \cite{he2024emilia}, a large-scale multilingual corpus containing 100,000 hours of paired audio and text that covers diverse accents and speaking styles. For ablation studies, we also use the 50,000-hour English LibriHeavy dataset \cite{kang2024libriheavy} and the 600-hour English LibriTTS dataset \cite{zen2019libritts}. Our evaluations are performed on the LibriSpeech-PC test-clean \cite{meister2023librispeech,chen2025f5} and the Seed-TTS test set \cite{anastassiou2024seed}. 

\vspace{-0.4em}
\subsection{Model Configuration}
\label{ssec:training_exp}
All VAE latents are pre-extracted prior to training the TTS model. Our base flow-matching model is trained for 800k updates with a batch size of 3,750 latent frames (approximately 0.67 hours of audio) using 8 RTX 5090 32GB GPUs over a span of 4 days.
AdamW optimizer is used with a peak learning rate of $1 \times 10^{-4}$, linear warmup for 1k updates, and linear decay. Gradient clipping is set to 1.0. We use $\eta=0.1$ for CTC loss.
The Exponential Moving Average (EMA) model is utilized for sampling. We allocate 18 DiT layers to the condition encoder and 4 layers to the velocity decoder. Additionally, the semantic aligner consists of 6 Transformer blocks.
The text input directly uses English alphabets for English text and pinyin representations for Chinese characters. During training, 70–100\% of audio latents are randomly masked for infilling. We extract speaker embeddings using a CAM++ model from 3D-Speaker\footnote{\url{https://www.modelscope.cn/models/iic/speech_campplus_sv_zh_en_16k-common_advanced}}. 
For CFG, the audio prompt and speaker embedding are jointly dropped with a probability of 0.3, while all conditioning is dropped of 0.2.

\vspace{-0.2cm}
\subsection{Baselines and Metrics}
\label{ssec:baselines}
\vspace{-0.1cm}
We compare ARCHI-TTS with state-of-the-art TTS models, including both AR and NAR models.
Evaluation is conducted under the cross-sentence task, focusing on word error rate (WER) and speaker similarity (SSIM). Following standard practices, WER is calculated using Whisper-large-v3 and paraformer-zh for English and Chinese transcription respectively, while SSIM is computed as the cosine similarity between speaker embeddings of synthesized and ground-truth speech, using a WavLM-large-based ECAPA-TDNN trained for speaker verification.

In addition, we perform Mean Opinion Score (MOS) evaluations, comparing ARCHI-TTS against the open-sourced F5-TTS \cite{chen2025f5} and the industrial-level CosyVoice2 \cite{du2024cosyvoice2} TTS model. We evaluate 10 samples from the Seed-TTS test set (5 English and 5 Chinese) using NMOS, SMOS, and CMOS metrics for naturalness, prompt similarity, and preference against ground truth, respectively. Sixteen raters participated in the evaluation.

\vspace{-0.2cm}
\subsection{Experimental Results}\label{sec:results}
\vspace{-0.1cm}
We report the average performance of our models based on 3 random seeds, as shown in Table~\ref{tab:librispeech}. Unless otherwise specified, we use a default of 32 NFE steps, a CFG strength of 4.0, and a timeshift of 3.0 for sampling. ARCHI-TTS demonstrates competitive performance on both the LibriSpeech-PC test-clean and the SeedTTS test set.

On LibriSpeech-PC testset, ARCHI-TTS achieves a WER of 1.98\% and SSIM of 0.70, demonstrating its ability to generate high-quality speech with strong speaker similarity. Given the low-token-rate VAE latents, our model can sample 10-second audio with a RTF of 0.21. On SeedTTS testset, ARCHI-TTS achieves a WER of 1.47\% and SSIM of 0.68 on English, and a WER of 1.42\% and SSIM of 0.70 on Chinese. This shows that ARCHI-TTS can still perform well on multi-lingual scenario within a single model. ARCHI-TTS archieves a WER of 2.14\% and a SSIM of 0.63 in 100k updates, indicating that the CTC Alignment contributes to faster convergence in WER. 
The results also indicate that our designs can achieve promising performance with limited resources. In the results of MOS evaluation shown in Tab~\ref{tab:mos}, ARCHI-TTS remains competitive but slightly lags behind other SOTA models. 

\begin{table}[t]
    \centering
    \caption{Results of MOS evaluation of ARCHI-TTS against other SOTA models on selected samples in SeedTTS testset.}
    \resizebox{0.7\linewidth}{!}{
    \begin{tabular}{l c c c}
        \toprule
        Model & NMOS & SMOS & CMOS \\
        \midrule
        Ground Truth & 3.72 & 3.59 & 0 \\
        \cdashline{1-4}\noalign{\vskip\belowrulesep}
        F5-TTS & 3.62 & \textbf{3.54} & -0.03 \\
        CosyVoice2 & 3.57 & 3.32 & 0.10 \\
        \cdashline{1-4}\noalign{\vskip\belowrulesep}
        ARCHI-TTS & 3.53 & \underline{3.48} & \underline{0.09} \\
        \bottomrule
    \end{tabular}}
    \label{tab:mos}
    \vspace{-0.6cm}
\end{table}

\begin{table}[t]
    \centering
    \caption{Ablation studies in model architecture on LibriSpeech-PC test-clean using identical batch size for 400k training updates.}
    \resizebox{0.9\linewidth}{!}{
    \begin{tabular}{l c c c}
        \toprule
        Spec. & Train Dataset & WER(\%)↓ & SSIM↑ \\
        \midrule
        ARCHI-TTS Small & \multirow{2}{*}{LibriTTS} & 2.88 & 0.55  \\
        w/o spk embed && 2.50 & 0.49 \\
        \cdashline{1-4}\noalign{\vskip\belowrulesep}
        ARCHI-TTS & \multirow{4}{*}{LibriHeavy} & \textbf{2.16}  &\textbf{0.71} \\
        w/o spk embed & & 2.48 & 0.62\\
        w/ sem. VQ & & 2.48 & \textbf{0.71}\\
        { }{ }{ }{ } codebook size$\times 2$ & & \textbf{2.15}  &\textbf{0.71}\\
        
        \bottomrule
    \end{tabular}
    }
    \label{tab:ablation}
\end{table}

\begin{figure}[t]
 \centering
 \includegraphics[width=\linewidth]{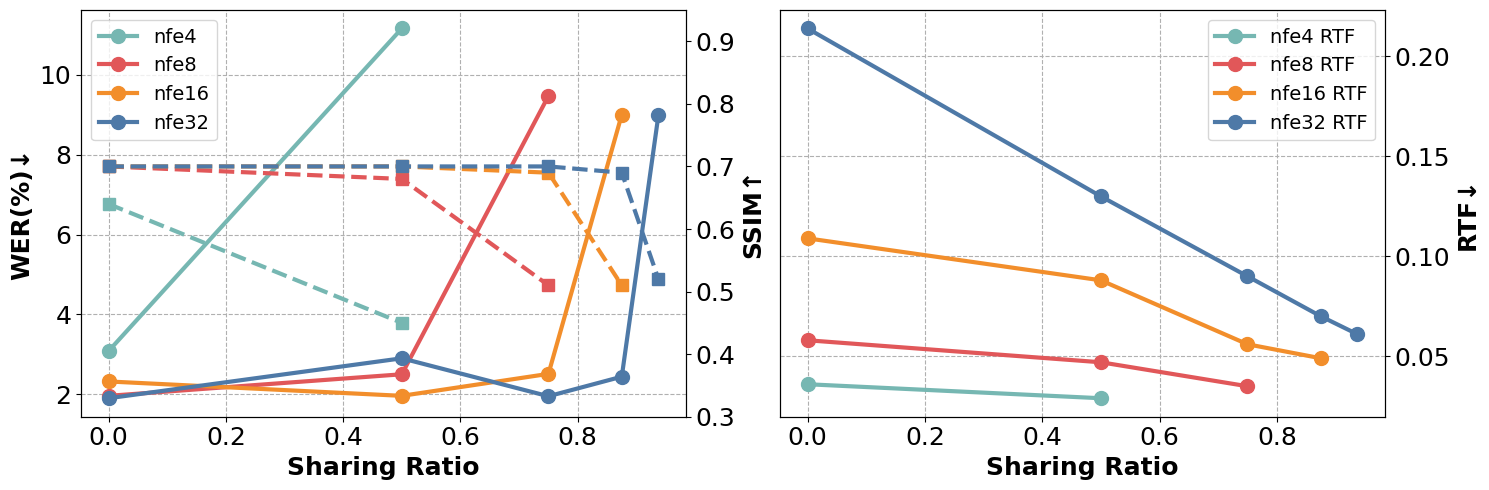}
 \vspace{-0.6cm}
 \caption{WER (solid line, left) and SSIM (dotted line, left), and RTF (right) across different sharing ratios.}
 \label{fig:K}
 \vspace{-0.6cm}
\end{figure}

To further evaluate the impact of our architectural choices, we perform ablation studies using small models trained on LibriTTS and base models trained on LibriHeavy. We conducted studies on adding Vector Quantization on semantic features as we found out it can lead to improvements in metrics. We focus on three aspects: (i) the effect of speaker embedding, (ii) the performance change introduced by vector quantization (VQ), and (iii) the influence of codebook size. Tab~\ref{tab:ablation} summarizes the results. From the results on small models, adding speaker embedding increases SSIM to 0.55 with a slight increase in WER. Scaling up model and data size without speaker embedding improves both metrics, while including speaker embedding further boosts SSIM with unchanged WER. These findings suggest that both larger datasets and speaker embedding contribute to improved prompt speaker similarity when using low token rate VAE latents as audio representations. In the base model, removing speaker embedding leads to a significant drop in SSIM, indicating the importance of speaker embedding in our system. It also implies that the commonly used mel-spectrograms contain richer speaker identity information than low-token-rate VAE latents which compress mostly the acoustic information. With VQ on semantic features, WER slightly increases. However, the WER is even lower than the original model with a doubled codebook size. This shows that a VQ-regularized semantic representation is beneficial to the model.

We can significantly reduce the computational cost of inference by sharing the conditioning hidden states across multiple flow steps. 
The results, shown in Fig.~\ref{fig:K}, reveal that a high sharing ratio significantly accelerates inference, though with some performance degradation. However, increasing the NFE steps can partially compensate for this performance drop. At an NFE of 32, ARCHI-TTS achieves a WER of 1.98\% and a SIM of 0.70, operating at an RTF of 0.09 with a 75\% sharing ratio, demonstrating both efficiency and robustness in low-latency synthesis without extra training required.

\vspace{-0.3cm}
\section{Conclusion}
\vspace{-0.2cm}
\label{sec:conclusion}
In this paper, we introduced ARCHI-TTS, a non-autoregressive text-to-speech model featuring a novel semantic aligner and a flow-matching decoder. Experimental results demonstrate that ARCHI-TTS achieves competitive performance over the SOTA models on the LibriSpeech-PC and Seed-TTS testset. 
Ablation studies confirmed the effectiveness of our core components, including the use of low-token-rate VAE latents and a training-free inference acceleration method that preserves speech synthesis quality. Future work will focus on sampling acceleration strategies in the decoder.

\vspace{-0.3cm}
\section{Acknowledgment}
\vspace{-0.2cm}
This project is partly funded by NSFC Young Scientists Fund (Category C) No. 62501512.

\vfill\pagebreak

\bibliographystyle{IEEEbib}
\bibliography{strings,refs}

\end{document}